\documentclass[12pt]{iopart}
\usepackage{epsfig}

\newcommand{\be}{\begin{equation}}
\newcommand{\ee}{\end{equation}}
\newcommand{\bea}{\begin{eqnarray}}
\newcommand{\eea}{\end{eqnarray}}
\newcommand{\pd}{\partial}
\newcommand{\bfig}{\begin{figure}[ht!]}
\newcommand{\efig}{\end{figure}}

\newcommand{\WP}[1]{}

\newcommand{\LB}[1]{\label{#1}}

\begin{document}

\title[Whirling Chaos in Convection]{Whirling Hexagons and Defect Chaos in
Hexagonal Non-Boussinesq Convection}

\author{Yuan-Nan Young\dag\, Hermann Riecke\ddag, and Werner Pesch$^+$}

\address{\dag\ Center for Turbulence Research, Stanford University,
Stanford, CA 94305,USA}

\address{\ddag\ Department of Engineering Sciences and Applied Mathematics,
Northwestern University, Evanston, IL 60208, USA}

\address{$^+$ Physikalisches Institut, Universit\"at Bayreuth, D-95440
Bayreuth, Germany}

\begin{abstract}

We study hexagon patterns in non-Boussinesq convection of a thin
rotating layer of water. For realistic parameters and boundary
conditions we identify various linear instabilities of the pattern.
We focus on the dynamics arising from an oscillatory side-band
instability that leads to a spatially disordered chaotic state
characterized by oscillating (whirling) hexagons. Using triangulation
we obtain the distribution functions for the number of pentagonal and
heptagonal convection cells. In contrast to the results found for 
defect chaos in the complex Ginzburg-Landau equation and in
inclined-layer convection, the distribution functions can show
deviations from a squared Poisson distribution that suggest
non-trivial correlations between the defects.

\end{abstract}



\maketitle

\section{Introduction}

Rayleigh-B\'enard convection has served as an excellent  paradigm to
study spatio-temporal chaos in pattern-forming systems.  For small
Prandtl numbers, for which large-scale flows driven by the curvature
of the roll pattern are important, quite complex dynamical roll
patterns exhibiting various types of defects like  spirals, targets,
dislocations, and disclinations are found \cite{MoBo93}. Another type
of spatio-temporally chaotic state is obtained if the system is
rotated about a vertical axis. Then the K\"uppers-Lortz instability 
\cite{KuLo69} leads to the formation of domains of rolls of different
orientation, which are separated by domain walls.
Due to the propagation of the domain walls or by the 
nucleation of defects in the  bulk of the pattern a persistent 
switching of the local roll wavevector of the different domains is induced \cite{HuPe98}. 

While striped planforms typically prevail in convection, hexagonal
cellular patterns  are observed at onset if the  temperature
variation across the layer is large and the variation of the material
parameters with the temperature becomes substantial. In this case the
Boussinesq approximation is invalid and a certain up-down reflection
symmetry is strongly broken \cite{Bu67}. Recently, results have been
presented also for spatio-temporal chaos based on a hexagonal rather
than a stripe-like planform. Within the framework of  Ginzburg-Landau
equations and of order-parameter equations of the  Swift-Hohenberg 
type the effect of rotation \cite{EcRi00,EcRi00a,YoRi03a} and of
large-scale flows \cite{YoRi02}  on hexagonal patterns has been
investigated.  Two interesting scenarios of spatio-temporal chaos
have been found that are both due to the rotation. One scenario
arises from the Hopf bifurcation to oscillating hexagons that
replaces the common steady transition from hexagons to rolls
\cite{Bu67} in the presence of rotation \cite{Sw84,So85}.  The
oscillations of the hexagons are described by the complex
Ginzburg-Landau equation \cite{ArKr02}.  If the transition  to the
oscillating hexagons occurs close to threshold the parameters of the
complex Ginzburg-Landau equation are generically in a regime in which
spatially homogeneous and spatially modulated oscillations (i.e.
traveling waves in the oscillation amplitude) are bistable with
spiral-dominated defect chaos \cite{EcRi00a}. For general initial
conditions, which induce the formation of spirals, the oscillations
of the hexagons are therefore expected to be spatio-temporally
chaotic.  The other scenario involves the steady-hexagon state below
its transition to oscillating hexagons. Within a weakly nonlinear 
Swift-Hohenberg framework a chaotic state  was found that is
maintained by  the nucleation of dislocations in the vicinity of
already existing penta-hepta defects. Due to this induced nucleation
the probability distribution function for the number of defects in
the pattern is substantially  broader than that obtained in
simulations of defect chaos in the complex Ginzburg-Landau equation
\cite{GiLe90} and in a model for $Ca^+$-waves \cite{FaBa99}, as well
as in experiments on electroconvection of nematic liquid crystals
\cite{ReRa89} and on  undulation chaos in inclined-layer convection
\cite{DaBo02,DaBo02a}.

Motivated by the interesting dynamics of defects in the oscillation
amplitude or in the hexagonal pattern itself that were obtained in
the Ginzburg-Landau and the Swift-Hohenberg models we investigate 
here the dynamics of hexagonal patterns in the presence of rotation
on the basis of the standard hydrodynamical description of 
non-Boussinesq convection  \cite{Ch61,Bu67,Bu89a,BoPe00}. Thus, we
use the Navier-Stokes equations coupled to the heat diffusion
equation (called NONBE henceforth, see below)  corresponding to
convection in a layer of water near its density maximum.   We perform
Galerkin stability calculations for the hexagonal pattern and direct
numerical simulations of the NONBE and find spatio-temporal chaos
that involves an oscillatory instability of the hexagons as well as
defects in the hexagonal background pattern. In contrast to the
defect chaos of the oscillating hexagons studied in the weakly
nonlinear regime \cite{EcRi00}, here the oscillatory instability does
not correspond to spatially homogeneous oscillations  but involves a
spatial modulation of the oscillation amplitude and it leads to the
nucleation of defects in the hexagonal background pattern.

\section{Basic Equations}

\LB{s:basic}

A horizontal fluid layer of
thickness $d$ that is heated from below and  cooled from
above and that is rotated about a vertical axis with angular
velocity $\omega \vec{\bf k}$ is described by the Navier-Stokes
equation for the momentum,
 \bea 
\pd_t (\rho  u_i)+\pd_j (\rho u_j u_i)&=&- \pd_ip -
\rho g\delta_{i3}+\nonumber \\ & &\pd_j \left(\nu \rho \,
(\pd_iu_j+\pd_ju_i)\right)+ 2 \rho \omega \epsilon_{ilm} k_l u_m, 
\LB{e:NS00} 
\eea 
the continuity equation, 
\be
\pd_t\rho+\pd_j(\rho u_j)=0, \LB{e:cont00} 
\ee 
and the heat equation,
\be \pd_tT+u_j\pd_jT=\frac{1}{\rho c_p}\pd_j(\lambda \pd_jT).
\LB{e:heat00} 
\ee
Here $\vec{u}=(u_1,u_2,u_3)$ is the fluid velocity, $T$ the
temperature, $\rho$ the  density, $p$ the pressure, $g$ the
acceleration of gravity, $\nu$ the viscosity, $\lambda$ the heat
conductivity, $c_p$ the specific heat of the fluid, and $\vec{\bf k}$
is the unit vector in the $z$-direction. The Kronecker delta is given
by $\delta_{ij}$  and $\epsilon_{ijk}$ is the unit antisymmetric
tensor of the rank 3. We assume the Einstein summation convention. We
have omitted  viscous heating, volume viscosity, and the centrifugal
force.  

In our analysis we take realistic rigid boundary conditions at the
top and bottom plates and keep the top and bottom temperatures fixed,
\bea 
\vec{u}&=&0  \mbox{ at }
z=\pm\frac{d}{2},\LB{e:bcrigid}\\ 
T&=&T_0+\frac{\Delta T}{2}  \mbox{
at } z=-\frac{d}{2},\LB{e:bcTbot}\\ 
T&=&T_0-\frac{\Delta T}{2} 
\mbox{ at } z=+\frac{d}{2}.\LB{e:bcTtop} 
\eea 

Here $T_0$ denotes the mean temperature and $\Delta T> 0$ is the
temperature difference across the layer. In the lateral directions we
chose periodic boundary conditions or introduce a suitable ramp in
the imposed temperature gradient (see below).  

In this paper we focus on weakly non-Boussinesq convection. We
therefore keep the temperature dependence of the various fluid
properties to leading order and expand them about the mean
temperature $T_0$ in the form \cite{Bu67}
\bea
\frac{\rho(T)}{\rho_0}&=&1-
\gamma_0 \frac{T-T_0}{\Delta T}(1+\gamma_1\frac{T-T_0}{\Delta T})+...\LB{e:NBrho}\\
\frac{\nu(T)}{\nu_0} &=& 1+\gamma_2 \frac{T-T_0}{\Delta T}+...\LB{e:NBnu}\\
\frac{\lambda(T)}{\lambda_0} &=& 1+\gamma_3\frac{T-T_0}{\Delta T}+...\LB{e:NBlambda}\\
\frac{c_p(T)}{c_{p0}}&=&1+\gamma_4\frac{T-T_0}{\Delta T}+...\LB{e:NBcp}
\eea
where $\rho_0$, $\nu_0$, $\lambda_0$, and $c_{p0}$ denote the values
of the respective quantities at the mean temperature $T_0$. The dots
denote higher-order terms to be neglected in the sequel. In line with
a clever experimental strategy we assume that $T_0$ is kept fixed
when changing the main control parameter $\Delta T$. Thus, the
quantities $\gamma_i /\Delta T , i = 0, 2, 3, 4$, which describe,
respectively, the slope of the density, viscosity, heat conductivity,
and heat capacity at $T_0$, are fixed as well. This implies that the
$\gamma_i$, which give the change of the respective fluid property
across the fluid layer, are linear in the main control parameter 
$\Delta T$ and therefore need to be 
adjusted with increasing $\Delta T$ (see below). 
The usual heat
expansion coefficient at $T = T_0$ is given by
$\alpha_0=\gamma_0/\Delta T$, but going beyond the Boussinesq
approximation also the curvature of $\rho (T) $ at $T_0$, which is
proportional to $\gamma_0 \gamma_1/{\Delta T ^2}$, comes into play.

Note that for constant fluid parameters the volume viscosity term and
the centrifugal force can be absorbed into the pressure term. This is
not true when the properties depend on the temperature. While the
$z$-dependence of the volume viscosity generates only a higher-order
correction, neglecting the centrifugal force is only  possible as
long as the rotation rates are not too large for a given system size
$L$,  i.e. as long as $\gamma_0 L\omega^2/g$ is small.  

To render the equations dimensionless we use the usual scales for the
length ($d$), time ($d^2/\kappa_0$), velocity ($\kappa_0/d$), 
pressure ($\nu_0 \kappa_0\rho_0/d^2$), and temperature  ($\nu_0
\kappa_0/\alpha_0 g d^3\equiv\Delta T/R$). Here we have introduced
the  heat diffusivity $\kappa_0=\lambda_0/\rho_0 c_{p0}$. This leads
to the usual dimensionless quantities like the Prandtl number
$Pr=\nu_0/\kappa_0$, the Rayleigh number $R=\alpha_0 \Delta T g
d^3/\nu_0 \kappa_0$, and the rotation rate $\Omega = \omega
d^2/\nu_0$.  Finally, we write the equations in terms of the
dimensionless momenta $v_i=\rho d u_i/\rho_0 \kappa_0$ instead of the
velocities. 

Since the fluid velocities are small compared to the sound velocity
we make the anelastic approximation \cite{Go69} (see also
\cite{Bu89a}) and neglect the time derivative in the continuity
equation (\ref{e:cont00}). This simplifies the computation
considerably since it reduces the number of evolution equations.
Furthermore, $v_i$ becomes a solenoidal field, which can be
represented in the standard poloidal-toroidal decomposition by two
velocity potentials \cite{Bu89a}, which automatically enforces the
mass conservation.

The conduction solution ($\vec v = 0$) of (\ref{e:heat00}) with
(\ref{e:bcTbot},\ref{e:bcTtop},\ref{e:NBrho},\ref{e:NBlambda},\ref{e:NBcp})
is given by
\bea
T_{cond}=T_0+
R\left(-z-\frac{\gamma_3}{2}(z^2-\frac{1}{4})+{\mathcal O}(\gamma_3^2)\right).\LB{e:cond}
\eea
We rewrite the temperature $T$ in terms of the deviation $\Theta$ from the
conductive profile (\ref{e:cond}) neglecting its contribution of ${\mathcal O}(\gamma_3^2)$,
\bea 
\Theta=T-T_{cond}=T-T_0-R\left(-z-\frac{\gamma_3}{2}(z^2-\frac{1}{4})\right). \LB{e:Tdev}
\eea

We then obtain as the final dimensionless equations
\bea
\frac{1}{Pr}\left(\pd_tv_i+v_j\pd_j\left(\frac{v_i}{\rho}\right)\right)&=&-\pd_i p 
+\delta_{i3}\left(1+\gamma_1(-2 z+\frac{\Theta}{R})\right)\Theta+\nonumber \\
&&+\pd_j\left[\nu\rho\left(\pd_i(\frac{v_j}{\rho})+\pd_j(\frac{v_i}{\rho})\right)\right]+
2 \Omega
\epsilon_{ij3}v_j, \LB{e:v}\\
\pd_jv_j&=&0, \LB{e:cont}\\
\pd_t\Theta+\frac{v_j}{\rho}\pd_j\Theta
& =&\frac{1}{\rho
c_p}\pd_j(\lambda\pd_j\Theta)-\gamma_3\pd_z\Theta-R\frac{v_z}{\rho}(1+\gamma_3z).\LB{e:T}
\eea
The dimensionless boundary conditions are
\bea
\vec{v}(x,y,z,t)=\Theta(x,y,z,t)=0  \mbox{ at } z= \pm \frac{1}{2}.\LB{e:bc}
\eea
The nondimensionalized  fluid parameters (\ref{e:NBrho})-(\ref{e:NBcp}) read now:
\bea
\rho(\Theta)&=&1-\gamma_0(-z+\frac{\Theta}{R}),\LB{e:rhoTh}\\
\nu(\Theta)&=& 1+\gamma_2(-z+\frac{\Theta}{R}),\LB{e:nuTh}\\
\lambda(\Theta)&=&1+\gamma_3(-z+\frac{\Theta}{R}),\LB{e:lambdaTh}\\
c_p(\Theta)&=&1+\gamma_4(-z+\frac{\Theta}{R}).\LB{e:cpTh}
\eea
We consider the non-Boussinesq effects to be weak and keep in all
material properties only the leading-order temperature dependence
beyond the Boussinesq approximation. Therefore the $\gamma_1$-term
appears explicitly in  (\ref{e:v}), while in all other terms  it
would constitute only a quadratic correction just like the terms
omitted in (\ref{e:NBrho})-(\ref{e:NBcp}). Correspondingly, we expand
the denominators in (\ref{e:v},\ref{e:T}) that contain material
properties  to leading order in the $\gamma_i$. In analogy to
\cite{Bu67} (eq.(6.7)), we further omit non-Boussinesq terms that
contain cubic nonlinearities in the amplitudes $v_i$ or $\Theta$, as
they arise from the expansion of the advection terms $v_j
\partial_j(v_i/\rho)$ and $(v_j/\rho)\partial_j \Theta$, once the
temperature-dependence of the density is taken into account.  Since we  will be
considering Rayleigh numbers up to twice the critical value, which
implies enhanced non-Boussinesq effects, these approximations may lead
to quantitative differences  compared to the fully non-Boussinesq
system, even though the temperature-dependence of the material
properties themselves may quite well be described by a linear (or
quadratic in the case of the density) approximation.

\section{Computational methods}

We solve the NONBE (\ref{e:v},\ref{e:cont},\ref{e:T}) with top and bottom
boundary conditions given by (\ref{e:bc}) and the material parameters by 
(\ref{e:rhoTh},\ref{e:nuTh},\ref{e:lambdaTh},\ref{e:cpTh}) numerically using
 a number of approaches. 

As usually, the stability properties of the patterns are determined
by Galerkin methods. We assume an infinite extension of the layer in
the lateral directions, which is captured by a Fourier expansion on a
hexagonal lattice. The Fourier wave vectors $\bf q$ are constructed
as linear combinations  of the hexagonal basis vectors ${\bf b}_1
=q(1,0)$ and  ${\bf b}_2 =q(1/2, \sqrt{3}/2)$ as  ${\bf q} = m {\bf
b}_1 + n {\bf b}_2$ with the integers $m$ and $n$ in the range
$|m|+|n| \le n_q$.  The largest wavenumber is then $n_q q$ and the
number of Fourier modes retained is given by $1+6\sum_{j=2}^{n_q}j$. 
Typically we used $n_q =3 $. The top and bottom boundary conditions
are satisfied by using appropriate combinations of trigonometric and
Chandrasekhar functions in $z$ \cite{Ch61,Bu67}. In most of the
computations we used $n_z=6$ modes for each field in
eq.(\ref{e:v},\ref{e:cont},\ref{e:T}). The linear analysis yields the
critical Rayleigh number $R_c(\gamma_i)$ as well as the critical
wavenumber $q_c(\gamma_i)$. In the large $Pr$-number regime, to be
considered in this paper, the bifurcation to convection is
stationary. Note that $q_c(\gamma_i)$, $R_c(\gamma_i)$ show
$O(\gamma_i^2)$  corrections compared to the standard Boussinesq
values $R_c = 1708$ and $q_c = 3.118$ for $\gamma_i = 0$.  

To investigate the nonlinear hexagon solutions, we start with the
standard weakly nonlinear analysis to determine the coefficients of
the three coupled amplitude equations for the modes making up the
hexagonal pattern. This allows us to get a first insight into the
K\"uppers-Lortz instability of the rolls, and for the weakly
non-Boussinesq case the transition from hexagons to rolls as well as
the transition to oscillating hexagons. To obtain the fully nonlinear
solutions requires the solution of a set of nonlinear algebraic
equations for the expansion coefficients with respect to the Galerkin
modes. This is achieved with a Newton solver for which the weakly
nonlinear solutions serve as convenient starting solutions. The fully
nonlinear solutions are then tested for amplitude and modulational
instabilities following the standard methods. Alternatively, we solve
the NONBE by direct numerical simulation to describe in general the
temporal evolution of the pattern. While this allows for a precise
check of the Galerkin results, its main purpose is to capture the
complex spatio-temporal dynamics resulting from the instabilities of
the hexagonal pattern, which are not accessible in the Galerkin
approach. The simulation code involves slight extensions of a
previous, well-proven version \cite{Pe96,DePe94a,BoPe00} to include
the effect of the non-Boussinesq effects. 

\section{Results}

Instead of extensive parameter studies, we concentrate in this paper 
on a specific, interesting scenario that should be experimentally
realizable. We focus our investigation on fluids with a fairly large
Prandtl number (water). For smaller Prandtl numbers our previous
results using a Swift-Hohenberg equation and also some initial test
runs using the Navier-Stokes equations indicate that rolls
\cite{Bu67}, which may be nucleated in defects or at the wall, take
over the hexagon pattern already for relatively small values of the
Rayleigh number. Furthermore, we have confined ourselves in this work
on one representative rotation rate ($\Omega=65$, see below).

To obtain strong non-Boussinesq effects the layer would have to be
taken quite thin. However, then the dimensionless rotation rates
obtained for a given physical rotation rate are relatively low.
Therefore we decided to consider water near its density maximum as
the convecting fluid. Strong non-Boussinesq effects at larger layer
thickness are also obtained with glycerin. However, since the
rotation rate is made dimensionless using the viscous diffusion time $d^2/\nu$ 
the extremely large viscosity $\nu$ of glycerin yields only small
dimensionless rotation rates. The second row in Table \ref{t:NBtable}
gives the non-Boussinesq parameters used in all computations in this
paper. For $\Omega=65$
they correspond to a critical temperature difference of $10^0$C
across the layer at a mean temperature of $12^o$C. Note that for a
given thickness of the fluid layer the critical Rayleigh number
increases significantly with the rotation rate. Therefore, the
coefficients $\gamma_i^{(c)}$, which give the non-Boussinesq effects
at onset, also increase strongly with the rotation rate. Consequently,
while in the absence of rotation the system is essentially Boussinesq
for the chosen layer thickness of $d=0.46cm$ (first line in table
\ref{t:NBtable}) it is strongly non-Boussinesq for $\Omega=65$.

\begin{table}
\begin{tabular}{|l|l|l|l|l|l|l|l|}\hline 
$T_0$ [$^oC$]& $\Delta T_c$ [$^oC$] & $Pr$ & $\gamma_0^{(c)}$ &
$\gamma_1^{(c)}$ & $\gamma_2^{(c)} $ & $\gamma_3^{(c)}$ &
$\gamma_4^{(c)}$ \\ \hline 
12 & 2.81 & 8.728 & 0.0003 & 0.1652 & -0.0765 & 0.0097 & -0.0011 \\ \hline 
12 & 10 & 8.728 & 0.0012 & 0.5888 & -0.2760 & 0.0345 & -0.0038 \\ \hline 
\end{tabular}  
\caption{
Values for the non-Boussinesq coefficients $\gamma_i^{(c)}$ and the Prandtl
number $Pr$ for water at onset as a function of the mean temperature and the temperature
difference. } \LB{t:NBtable} 
\end{table} 
 
\begin{table}
\begin{tabular}{|l|l|l|l|l|l|l|l|l|}\hline 
$\Omega$  &  $T_0$ [$^oC$]& $\Delta T_c$ [$^oC$] &
$R_{c0}$ & $R_{c\gamma}$ & $q_{c0}$& $q_{c\gamma}$& $\tau_{th} [s]$ & $\omega [Hz]$ \\ \hline 
0  & 12 & 2.81 & 1708 & 1706 & 3.12 & 3.12& 150.7 & 0     \\ \hline 
65 & 12 & 10   & 6121 & 6080 & 5.27 & 5.39& 64.6 & 0.600 \\ \hline 
\end{tabular} 
\caption{Dependence of the critical Rayleigh
number and of the critical wavenumber on the rotation rate for a layer
thickness of $d=0.46cm$. The critical Rayleigh numbers in the Boussinesq approximation
($\gamma_i=0$) and for $\gamma_i=\gamma_i(\Delta T= \Delta T_c)$ are denoted
by $R_{c0}$ and $R_{c\gamma}$, respectively.
Similarly for the critical wavenumber $q_{c0,\gamma}$. 
The physical rotation rate for this cell thickness is given by $\omega$ and
the vertical thermal diffusion time by $\tau_{th}$. } 
\LB{t:crittable} 
\end{table}

We first determine numerically the linear stability of the stationary
hexagon patterns as a function of the wavenumber and the control
parameter $\epsilon=(R-R_c(\gamma_i))/R_c(\gamma_i)$ for $\Omega=65$
and the corresponding non-Boussinesq coefficients $\gamma_i^{(c)}$ as
given in table \ref{t:NBtable}. As indicated before, the $\gamma_i$
are linear in the temperature difference $\Delta T$ and are therefore
given by $\gamma_i=\gamma_i^{(c)}\,(1+\epsilon)$. All results were
obtained with $n_q=3$ and $n_z=6$.  In the parameter regime
investigated these cut-off parameters are sufficient. As shown in
Fig.\ref{f:stabi} the hexagon pattern can undergo a variety of linear
instabilities in the nonlinear regime. Close to threshold the
dominant instability is a long-wave instability, i.e. its growth rate
vanishes for vanishing modulation  wave number $\beta$ (Floquet
parameter) but grows quadratically with $\beta$ for small $\beta$.
Very close to onset it can be steady (black circles). For slightly
larger $\epsilon$, however, it becomes oscillatory (red squares).  As
the control parameter is increased a steady instability at finite
Floquet parameter limits the stability region of the hexagons for
large wavenumbers (blue triangles). Increasing the control parameter,
on the low-wavenumber side the stability region becomes limited by an
oscillatory instability that sets in with a finite Floquet parameter
(green diamonds). For non-Boussinesq coefficients that differ
slightly from those of table \ref{t:NBtable} (by ${\mathcal O}(20\%
)$) an oscillatory instability with vanishing Floquet parameter
determines the stability limit for low $q$ and $\epsilon \approx 1$.
For the parameters in table \ref{t:NBtable} it is, however, preempted
by the Hopf bifurcation at finite modulation wavenumber. 

The spatially homogeneous and the short-wave oscillatory
instabilities are similar to the oscillatory instability of hexagons
that is found within the framework of three coupled Ginzburg-Landau
equations \cite{Sw84,So85,EcRi00,EcRi00a}. There it replaces the
steady instability of hexagons that in the absence of rotation leads
to an unstable mixed-mode solution. With rotation it turns into a
supercritical Hopf bifurcation at which the oscillating-hexagon
solution branches off the stationary hexagons.

\bfig \centerline{
\epsfxsize=10.0cm\epsfbox{Pictures/stabilimitsnew.eps} } 
\caption{Stability limits of stationary hexagon pattern for the
parameters given in Table \ref{t:NBtable} ($n_q=3$, $n_z=6$, except
for dark green diamond, which is for $n_q=5$, $n_z=8$): long-wave
steady (black circle), long-wave oscillatory  (red squares),
oscillatory short-wave instability (green diamonds),  steady
short-wave instability (blue triangles).} \LB{f:stabi}  
\efig  

Numerical simulations support the expectation that the modes
corresponding to the Hopf bifurcation  are similar to the oscillating
hexagons. In a small computational domain of length $L=2\cdot
2\pi/q$, which contains only four convection cells, the oscillatory
side-band instability is suppressed and the homogeneous Hopf
bifurcation arises when the parameters are chosen sufficiently far
beyond the stability limit corresponding to the short-wave
oscillatory instability (green diamonds in Fig.\ref{f:stabi}). This
instability leads to an elliptic deformation of the  convection cells
which oscillates in time, giving the cells the appearance as
if they were rotating or in fact whirling. In sufficiently large
domains the side-band instability leads to a spatial modulation of
these oscillations as shown in Fig.\ref{f:destab}a ($L=8\cdot 2\pi/q$
with $64\times 64$ Fourier modes). 

\bfig \centerline{
\epsfxsize=5.0cm\epsfbox{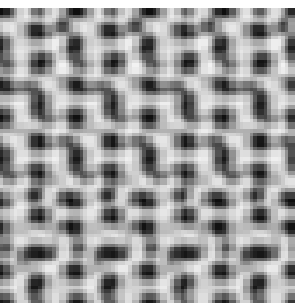}
\epsfxsize=5.0cm\epsfbox{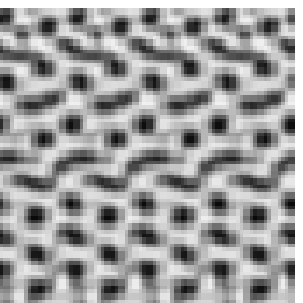}
\epsfxsize=5.0cm\epsfbox{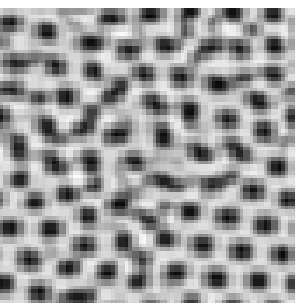}
}   
\caption{
Snapshots of the pattern for $\epsilon=1.0$, $L=8\times 2\pi/q$ with
$q=4.5$ (see also movie {\tt sn76.T.mpg}). Initial growth of modulational 
instability at $t=7.5$ (left panel) and $t=7.9$ (middle panel).  
Disordered state during an intermediate phase at $t=51.3$
(right panel). Black denotes hotter rising fluid.
}  \LB{f:destab}   
\efig 

As the instability develops and the oscillation amplitude grows the
regular spatial structure is destroyed and defects arise in the
pattern as pictured in Fig.\ref{f:destab}c. This temporal evolution
is illustrated in the movie {\tt sn76.T.mpg}. While the apparently
chaotic dynamics appear to be reaching a statistically steady state,
they do, however, not persist and around $t=130$ a relative rapid
transition to an ordered, stationary hexagon pattern occurs. It has a
slightly larger wave number $q$, which is in the stable regime (cf.
Fig. \ref{f:stabi}).  During this chaotic transient the overall
disordered pattern intermittently exhibits relatively large domains
of ordered hexagonal patterns. The duration of the transient
increases with $\epsilon$ as illustrated in Fig.\ref{f:transient},
reflecting presumably the narrowing of the band of stable wavenumbers
with increasing $\epsilon$.  To monitor the activity of the pattern
we consider as a representative measure the temperature field  in the
mid plane of the layer and plot the `whirling' activity ${\mathcal
W}$  given by the $L_1$-norm of the time derivative of its Fourier
modes \footnote{Since the temperature field is dominated by the
linear eigenvector with the maximal  growth rate this allows to
reduce the computational overhead.} normalized by the $L_1$-norm of
the Fourier modes themselves. As $\epsilon$ is increased the
fluctuations in this quantity decrease, which is expected to reduce
the probability of excursions to one of the linearly
stable stationary states, and eventually for $\epsilon=1.1$ our
simulations did not end up in a stationary pattern for times as large
as $t=400$.

\bfig \centerline{
\epsfxsize=8.0cm\epsfbox{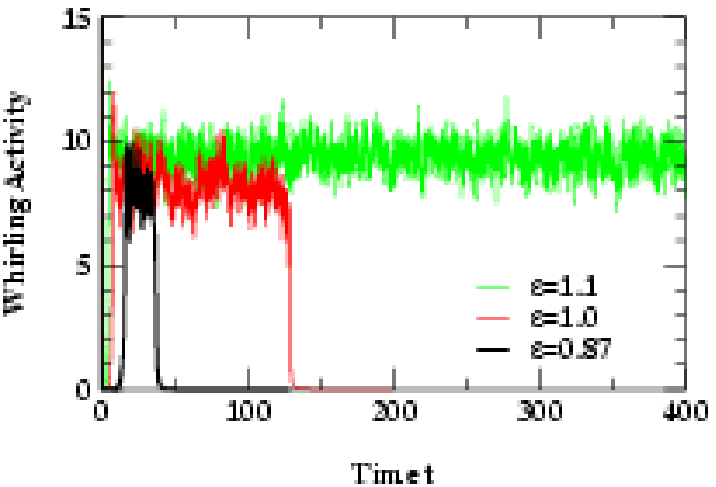}
 }   
\caption{Dependence of the duration of the transient on $\epsilon$ for
$L=8\times 2\pi/q$ with $q=4.5$ as quantified by the 
whirling activity ${\mathcal W}$ (see text). In physical units $t=400$ 
corresponds to about 7 hours.
}  \LB{f:transient}   
\efig 

The duration of the transients is presumably related to the size of
the system. We therefore performed also simulations in larger systems
of size $16\cdot 2\pi/q$ ($128\times128$ Fourier modes). For this
system size the dynamic state persists for quite a long time
($t_{max}\approx 300 \tau_{th} \approx 5 hrs $) even for control
parameters as low as $\epsilon=0.87$  (compare with the transients
for the smaller system shown in  Fig.\ref{f:transient}). This
increased duration of the transient is due to the longer time it
takes for the defects in the pattern to annihilate. For
$\epsilon=1.0$, a snapshot of the pattern and of the corresponding
temporal derivative is shown in Fig.\ref{f:large}a,b. The latter
makes the spatially intermittent behavior of the dynamics more
apparent. The dynamics are illustrated in the two corresponding
movies ({\tt sn79h.T.mpg, sn79h.D.mpg}).

\bfig \centerline{
\epsfxsize=5.5cm\epsfbox{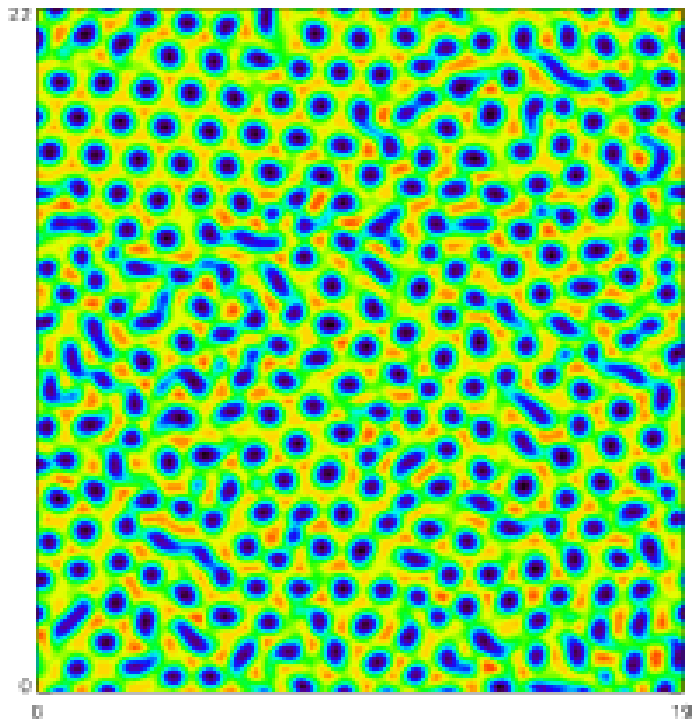} 
\epsfxsize=5.5cm\epsfbox{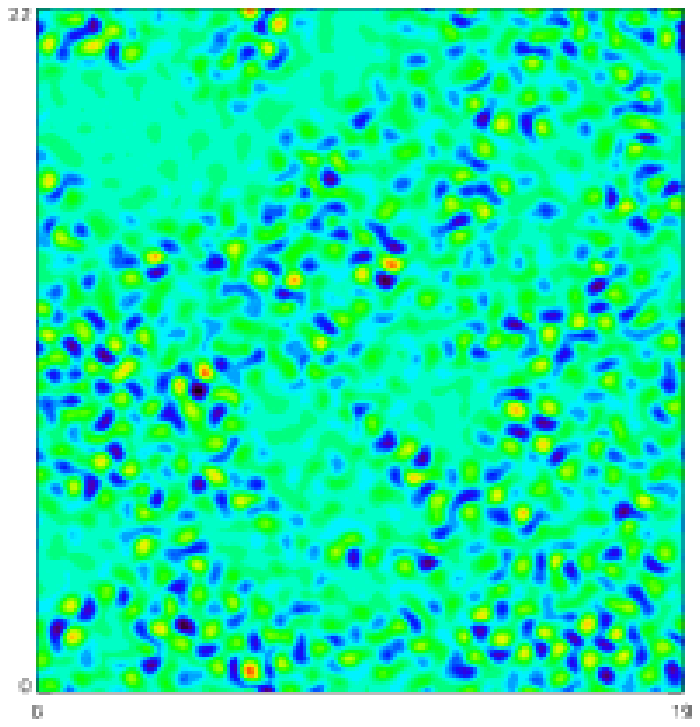} 
\epsfxsize=5.5cm\epsfbox{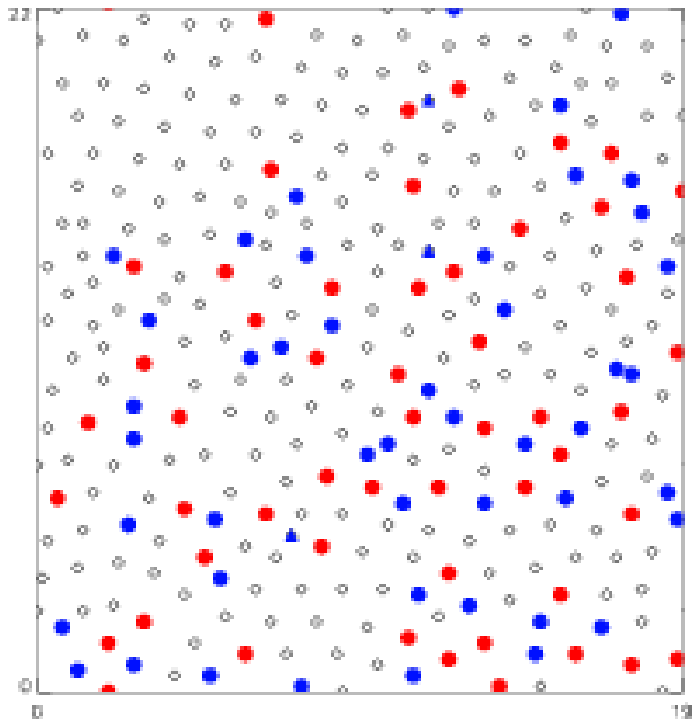} 
}   
\caption{Disordered state for $\epsilon=1.0$ and $L=16\times 2\pi/q$
with $q=4.5$. 
Snapshot of the pattern (blue denotes hotter rising fluid); 
see also movie {\tt sn79h.T.mpg} (left panel). Middle panel shows 
the difference between two successive snapshots (separated by $\Delta
t=0.1$) corresponding to the pattern shown in 
left panel (see also movie {\tt sn79h.D.mpg}). Blue (red) denotes locations 
with growing (decreasing) temperature.  
Right panel: identification of the minima with lattice nodes. 
Blue (red) circles have 5 (7) nearest neighbors. 
Up- (down-) triangles have 4 (8) neighbors.
}  \LB{f:large}   
\efig 

For the characterization of weakly disordered hexagon patterns it is often
advantageous to locate defects in the pattern by demodulating the
pattern using the (three) main wavevectors comprising the pattern
(e.g. \cite{YoRi03,YoRi03a}). This allows the
identification of the penta-hepta defects that are characteristic of
weakly disordered hexagon patterns and which consist of correlated
convection cells with five and seven neighbors, respectively. Obviously, 
this approach requires that the Fourier spectrum of the pattern
exhibits well-defined peaks at the three hexagon modes.

In the strongly disordered patterns obtained in our simulations the
spectrum does not exhibit clear peaks. Demodulation is therefore not
possible and other techniques for the analysis of the patterns have
to be brought to bear. We use a triangulation of the pattern to
locate the defects. To that end we map the convection pattern to a
point lattice by identifying the local minima in the temperature with
nodes in a lattice. To avoid spurious minima we require that the
minima fall below a certain threshold. The coordination number of
each node is then obtained using a Delaunay triangulation, which is
the unique triangulation obtained by the requirement that the circle
through the corners of each triangle do not contain any other node of
the pattern. We use the  program {\tt triangle} \cite{Sh02} to
perform the triangulation. Figure \ref{f:large}c shows the lattice 
of the snapshot shown in Fig.\ref{f:large}a with the local
coordination numbers obtained from the triangulation.  Black diamonds
indicate nodes with six neighbors, i.e. the center of a hexagonal
cell. Nodes with seven  (i.e. heptagons) and five neighbors
(pentagons) are marked with  red and blue circles, respectively,
while up- and down-triangles denote nodes with four and eight
neighbors, respectively. For less disordered patterns  pentagons and
heptagons are often strongly correlated and can be associated with
each other in pairs. These pairs correspond then to penta-hepta
defects and can be identified  with pairs of topological dislocation
defects in two of the three modes that make up the hexagonal pattern
(see e.g. Fig.25 in \cite{BoPe00}). Isolated pentagons or heptagons,
however, do not correspond to single dislocations.

In previous analyses of defect-dominated spatio-temporal chaos it was
found that the distribution function of the number of defects  in the
patterns gives some insight into the dynamics of the system. More
specifically, in the defect chaos obtained in the  complex
Ginzburg-Landau equation \cite{GiLe90}, in electroconvection of
nematic liquid crystals \cite{ReRa89}, and in undulation chaos in
inclined-layer convection \cite{DaBo02,DaBo02a} the statistics were
found to be close to  a Poisson-type distribution. Such a
distribution arises if defects of opposite topological charge are
created randomly in pairs with a fixed probability and are
annihilated with a rate that is proportional to the square of the
density of the defects. The Poisson statistics therefore suggest
that the defects move like uncorrelated random walkers. In contrast,
for the penta-hepta defect chaos that was found in hexagon patterns
in the presence of rotation  within the framework of a
Swift-Hohenberg-type equation the distribution  function was found to
be considerably broader \cite{YoRi03a}. The origin of this broadening
was identified to be the induced nucleation of dislocations by
penta-hepta defects through which also the creation rate depends on
the defect density  \cite{YoRi03a,CoNe02}.

\bfig  \centerline{
\epsfxsize=8.0cm\epsfbox{Pictures/sn79.57stats.eps} 
\epsfxsize=8.0cm\epsfbox{Pictures/sn79.48stats.eps} }  
\caption{a)
Probability distribution function for the number of pentagons (blue)
and heptagons (red) for $\epsilon=1$ and corresponding best fit to
the Poisson distribution  (\ref{e:poisson}) (dashed lines).\\ b)
Probability distribution function for the number of rectangles
(blue), octagons (red), and for the difference (${\mathcal
P}_{n_5-n_7}$)  between the number of pentagons and heptagons (solid
line). Parameters as in Fig.\ref{f:large}.} \LB{f:pdf65} 
\efig

Using the triangulation procedure, we have measured the distribution
functions for the number of pentagons and heptagons.  For the state
corresponding to Fig.\ref{f:large} they are presented  in
Fig.\ref{f:pdf65}a,b, where the blue circles and red triangles
denote the distribution functions for pentagons and heptagons,
respectively. As mentioned previously, in strongly disordered
patterns pentagons and  hexagons are not necessarily bound in
penta-hepta defects. Correspondingly, the number of pentagons does
not always agree with the number of heptagons. The distribution
function for their difference is given by the dashed line in Figure
\ref{f:pdf65}b along with the distribution function for rectangles
(blue circles) and octagons (red triangles).   

As a first step it appears reasonable to assume that the dynamics of
the pentagons and heptagons are captured by a kinetic model that is
analogous to the description of dislocations \cite{GiLe90}. For
periodic boundary conditions \cite{DaBo02} this leads then to the
standard Poisson distribution for each species,    \bea     
{\mathcal P}(N)=\frac{\alpha^N}{I(2\sqrt{\alpha})\Gamma(1+N)^2},  
\LB{e:poisson}     \eea       
where $\alpha\equiv <N^2>$ and
$I(2\sqrt{\alpha})$ is the modified Bessel function.  In some
simulations we observe that a given whirling cell creates a new
convection cell nearby, which implies the creation of defects. The
newly created cell then often merges with the whirling cell.
Conversely, defects in the pattern induce whirling activity. This
mutual nonlinear enhancement of defects and whirling activity
suggests that the existence of defects may lead to the creation of
further defects, somewhat similar to the induced nucleation in
penta-hepta defect chaos \cite{YoRi03a}. As shown by the dashed lines
in Fig.\ref{f:pdf65}a, for the state of Fig.\ref{f:large} the Poisson
distribution (\ref{e:poisson}) does not fit the pentagon and the
heptagon distributions very well. The distributions therefore suggest
that the creation rates for the pentagons and heptagons are not
independent of their respective densities. In some preliminary
parameter studies involving  non-Boussinesq coefficients that may not
correspond to a realistic convection cell with water as fluid, we
have found distributions that are fitted quite well by the Poisson
distribution and others that are even broader than the distributions
obtained in Fig.\ref{f:pdf65}. This suggests that a mechanism akin to
induced nucleation may be operative in this system, with its
significance depending on the specific parameter values
\cite{YoRiunpub}. A study of the dependence  of the distribution on
the physical parameters of the fluid is beyond the scope of the
present paper.

Fig.\ref{f:large}b shows that in the disordered state not all cells
are active and at times sizable patches of relatively ordered
quiescent domains arise. To characterize this intermittent behavior
we consider the distribution function for the time derivative of the
temperature as obtained from all snapshots of the type shown in
Fig.\ref{f:large}b. The logarithmic scale of Fig.\ref{f:pdfdTdt}
reveals an exponential decay of the distribution function  for
$\epsilon=1.0$. This exponential dependence is consistent with the
visually apparent spatio-temporal intermittency of the activity of
the pattern. The pronounced peak  exhibited for $\epsilon=0.87$
represents the large spatial domains in which the hexagons are
ordered and quiescent. 

\bfig  
\centerline{
\epsfxsize=9.0cm\epsfbox{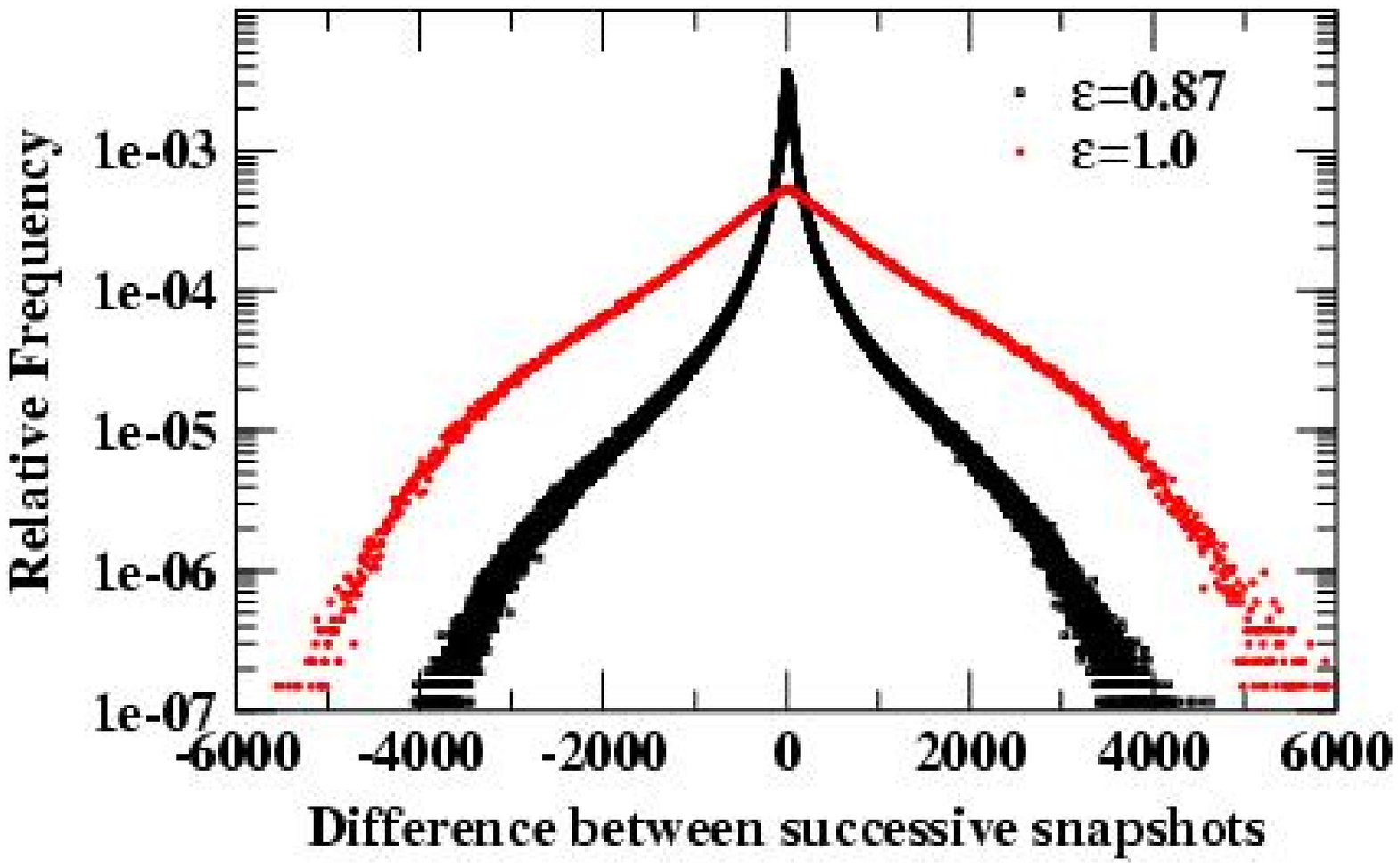} 
} 
\caption{  Probability distribution function for the difference
between successive snapshots ($\Delta t =0.1$) for 
$L=16\times 2\pi/8$ and $q=4.5$.} 
\LB{f:pdfdTdt} 
\efig

In all simulations so far we have employed periodic boundary
conditions. In experiments boundary effects may strongly perturb the
patterns. We therefore have also performed some simulations in which
we mimic a circular container by applying a strong radial subcritical
ramp in the Rayleigh number that suppresses any convection outside a
certain radius. In previous simulations \cite{BoPe00} this was seen
to provide a reasonable idea of the impact of boundaries.
Experimentally, it has been found that boundaries tend to induce
defects and may also promote the formation of rolls, which could
invade the system and replace the hexagonal pattern \cite{Ahpriv}.
For the parameter regimes investigated here we find that no patches
of rolls invading from the side replace the hexagons. Instead the
precession of the cells near the boundary and the associated
formation of defects triggers persistent dynamics in the whole
convection cell even for parameter values for which in the absence of
the boundaries the pattern eventually becomes ordered and stationary.
Fig.\ref{f:circle}a shows a snapshot for $\epsilon=0.87$ with the
associated temporal evolution displayed in the movie {\tt
sn74.c.T.mpg}. A snapshot for a larger system size and
$\epsilon=1.0$  is shown in fig.\ref{f:circle}b (movie {\tt
sn79h.c.T.mpg}). We therefore expect that the defect-dominated
spatio-temporal chaos will also be accessible in experiments.  

\bfig 
\centerline{
\epsfxsize=6.0cm\epsfbox{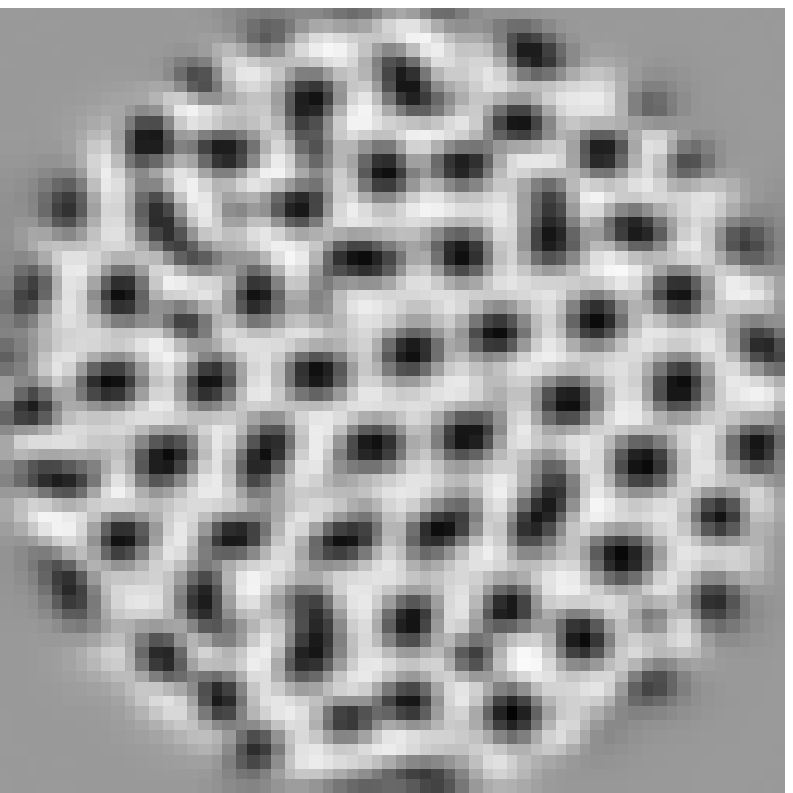} 
\epsfxsize=6.0cm\epsfbox{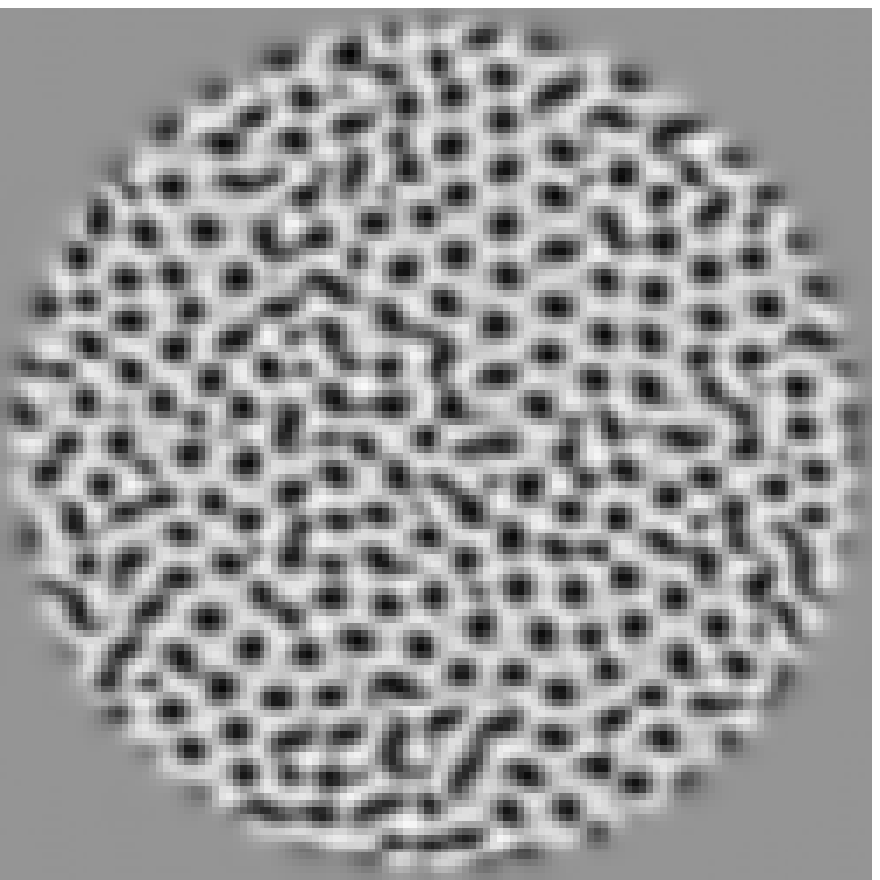}
} 
\caption{Snapshot of defect-dominated pattern in the presence of a radial
ramp mimicking a circular container. a) $\epsilon=0.87$, $L=8\cdot
2\pi/q$ with $q=4.5$. b) $\epsilon=1.0$, $L=16\cdot
2\pi/q$ with $q=4.5$. Temporal evolution shown in movies {\tt
sn74.c.T.mpg} and {\tt sn79h.c.T.mpg}.)
}
\LB{f:circle}
\efig

We have not investigated in detail the nonlinear evolution
ensuing from the steady instability at finite wavenumber that limits
the wavenumber band on the high-$q$ side. In preliminary simulations
the instability led to the formation of defects in the pattern, which
eventually, however, disappeared yielding a stationary, regular
hexagonal pattern.

\section{Conclusion}

In this paper we have considered hexagonal non-Boussinesq convection focusing
on rotating systems, in which the chiral symmetry is broken.  
The main motivation for this work stems from the interesting chaotic states that
had been identified in previous investigations of the stability 
of hexagonal patterns and of the dynamics resulting from their instabilities. 
The chaotic states were associated
with the secondary instability to oscillating hexagons
\cite{EcRi00,EcRi00a}  and with the induced nucleation of
dislocations \cite{YoRi03,YoRi03a,CoNe02}, respectively. These
results were obtained within order-parameter and Ginzburg-Landau
models. To make closer contact with experimental systems we have investigated in this
paper the effect of rotation on realistic hexagonal convection in a
thin layer of water in a temperature regime near the anomalous
density maximum.  

Using a numerical stability analysis and direct numerical simulations of
the Navier-Stokes equations we have identified an oscillatory side-band
instability, which leads to spatially modulated whirling hexagons that
in turn become disordered in space and exhibit irregular persistent
dynamics, which coexists with linearly stable regular hexagon patterns. 

Coexistence of spatio-temporal chaos with ordered patterns has been
observed previously in roll convection at low Prandtl numbers
\cite{MoBo93,CaEg97}. There the spiral defect chaos is sustained by the
mean flow, which compresses  the rolls so as to push them locally across
their stability limit with respect  to the skew-varicose instability,
which then leads to the formation of  defects \cite{EgMe98}. This
mechanism for sustaining dynamics is reminiscent of that observed 
earlier in a  much smaller systems \cite{Cr89,Cr89a}. 

The method employed in \cite{EgMe98} to analyze the disordered
pattern, which extracts the local wavenumber and orientation of the
rolls, is not applicable to the  strongly disordered hexagonal
patterns found in our simulations. We therefore use Delaunay
triangulation to measure the distribution functions for the number of
pentagons and heptagons. Both show deviations from a squared Poisson
distribution, which suggests that their dynamics (including their
creation and annihilation) exhibit correlations that are absent in
the dynamics of dislocations in the complex Ginzburg-Landau equation
\cite{GiLe90} or in inclined-layer convection \cite{DaBo02,DaBo02a}.
It is, however, not quite clear whether a kinetic description of the
pentagons and heptagons similar to that of dislocations or of
penta-hepta defects \cite{GiLe90,DaBo02,YoRi03a} is in fact
appropriate. To characterize the intermittent appearance of ordered
domains in the chaotic states we have determined the distribution
function for the temporal derivative of the temperature field. It
exhibits an exponential decay reflecting the spatio-temporal
intermittency. For smaller Rayleigh number a pronounced peak appears,
which reflects the large ordered, quiescent domains.

In this paper we have focused on one set of fluid parameters  (cf.
table~\ref{t:NBtable}). For these values the oscillatory instability
with finite modulation wavenumber preempts a spatially homogeneous
oscillatory instability. We found, however, that the situation can be
reversed if the non-Boussinesq parameters are changed only slightly.
In these cases, which may not correspond to a realistic convection
cell with water as fluid, we have found defect-chaotic states with
distribution functions that are significantly wider than the squared
Poisson distribution suggesting again a nonlinear interaction between
the whirling of the hexagons and the formation of defects, as well as 
distribution functions that are well described by the squared Poisson
distribution. In addition, we have identified a state in which the
whirling is spatially and temporally intermittent in the form of
bursts while the underlying hexagonal planform has essentially no
defects \cite{YoRiunpub}. 

\ack We thank G. Ahlers for providing us with the software to
determine the necessary fluid parameters.  We gratefully acknowledge
support by grants from the Department of Energy (DE-FG02-92ER14303),
NASA (NAG3-2113), and NSF (DMS-9804673).  YY acknowledges computation
support from the Argonne National Labs and the DOE-funded ASCI/FLASH
Center at the University of Chicago. The codes used for the
computations are extensions of codes that were developed over the
years at the Universit\"at  Bayreuth with contributions by W. Decker
and A. Tschammer as well as V. Moroz.

\vspace{1cm}


\end{document}